\documentclass[aps,pre,reprint, amsmath, amssymb,superscriptaddress]{revtex4-1}

\usepackage{bm}
\usepackage[retainorgcmds]{IEEEtrantools}
\usepackage{graphicx}
\usepackage{mathrsfs}
\usepackage{amsmath}
\usepackage{amssymb}
\usepackage{color}
\usepackage{amsfonts}
\usepackage{times,txfonts}
\usepackage{nicefrac}

\newcommand{\ud}{\,\mathrm{d}}

\DeclareMathOperator{\tr}{tr}

\newcommand{\trans}{^{\text{T}}}

\begin{document}

\title{Analysis of the conditional mutual information in ballistic and diffusive non-equilibrium steady-states}
\date{\today}
\author{William T. B. Malouf}
\affiliation{Instituto de F\'isica da Universidade de S\~ao Paulo,  05314-970 S\~ao Paulo, Brazil}
\affiliation{School of Mathematical Sciences, University of Nottingham, University Park, Nottingham NG7 2RD, United Kingdom}
\author{John Goold}
\affiliation{School of Physics, Trinity College Dublin, Dublin 2, Ireland}
\author{Gerardo Adesso}
\affiliation{School of Mathematical Sciences, University of Nottingham, University Park, Nottingham NG7 2RD, United Kingdom}
\author{Gabriel T. Landi}
\email{gtlandi@if.usp.br}
\affiliation{Instituto de F\'isica da Universidade de S\~ao Paulo,  05314-970 S\~ao Paulo, Brazil}

\begin{abstract}

The conditional mutual information (CMI) $\mathcal{I}(A\! : \! C|B)$ quantifies the amount of correlations shared between $A$ and $C$ \emph{given} $B$. 
It therefore functions as a more general quantifier of bipartite correlations in  multipartite scenarios, playing  an important role in the theory of quantum Markov chains. 
In this paper we carry out a detailed study on the behavior of the CMI in non-equilibrium steady-states (NESS) of a quantum chain placed between two baths  at different temperatures. 
These results are used to shed light on the mechanisms behind ballistic and diffusive transport regimes and how they affect correlations between different parts of a chain.
We carry our study for the specific case of a 1D bosonic chain subject to local Lindblad dissipators at the boundaries. 
In addition, the chain is also subject to  self-consistent reservoirs at each site, which are used to tune the transport between ballistic and diffusive. 
As a result, we find that the CMI is independent of the chain size $L$ in the ballistic regime, but decays algebraically with $L$ in the diffusive case. 
Finally, we also show how this scaling can be used to discuss the notion of local thermalization in non-equilibrium steady-states.

\end{abstract}
\maketitle{}

%
%
%
%

\section{Introduction}
When a system is coupled to two reservoirs kept at different temperatures, it  eventually reaches a non-equilibrium steady-state (NESS),  characterized by the existence of a   heat current flowing from the hot to the cold bath.
Substantial effort has been dedicated over the past decades in furthering our general understanding of NESSs. 
For instance, important insights have been gained on  the microscopic mechanisms responsible for the emergence of Fourier's law \cite{Rieder1967,Bolsterli1970,Prosen1999,Aoki2000,Aoki2000a,Saito2003a,Bernardin2005,Asadian2013,Karevski2009,Znidaric2011b,Znidaric2010a,Landi2013a,Landi2014a}, the development of steady-state fluctuation theorems \cite{Lebowitz1999,Saito2007,Kundu2011,Panasyuk2012,Nicolin2011} and the description of NESSs  supporting the transport of more general types of excitations, such as magnetization currents \cite{Znidaric2011b,Znidaric2010a,Karevski2009,Znidaric2011,popkov2,Mendoza-Arenas2014a,Landi2014b,Landi2015a,Carollo2018} and even radiation squeezing \cite{Clark2016,Manzano2016}.

Informational aspects, however, have been much less explored. 
Excitations traveling through a chain also carry information \cite{Calabrese2005,Alba2017}, so that currents ultimately change the way different parts of a chain become correlated. 
A thorough understanding of how correlations between different part of the system behave in the NESS could provide valuable information on the multipartite structure of non-equilibrium states. 
Studies on this topic, however, are still scarce \cite{Zanoci2016,Mahajan2016,James2018,Horowitz2014a,Ptaszynski2019},
despite already being within reach of platforms such as ultra-cold atoms \cite{Brantut2013}, trapped ions \cite{Bermudez2013} and opto-mechanical systems \cite{Clark2016,Ockeloen-Korppi2018}.

One of the key questions in studies of the NESS, is what are the ingredients required to produce different kinds of transport regimes.
Integrable systems usually tend to NESSs with a ballistic heat flow, where the heat current $J$ is independent of the system size $L$. 
Non-integrable systems, on the other hand, tend in general to have currents scaling as $J\sim 1/L^\alpha$, where $\alpha$ is an exponent that has the value $\alpha = 1$ in the diffusive case~\cite{Dhar2008,Conf2011}.
It is then natural to enquire how the correlations between parts of a chain behave in different transport regimes.

\begin{figure}[!t]
\centering
\includegraphics[width=0.4\textwidth]{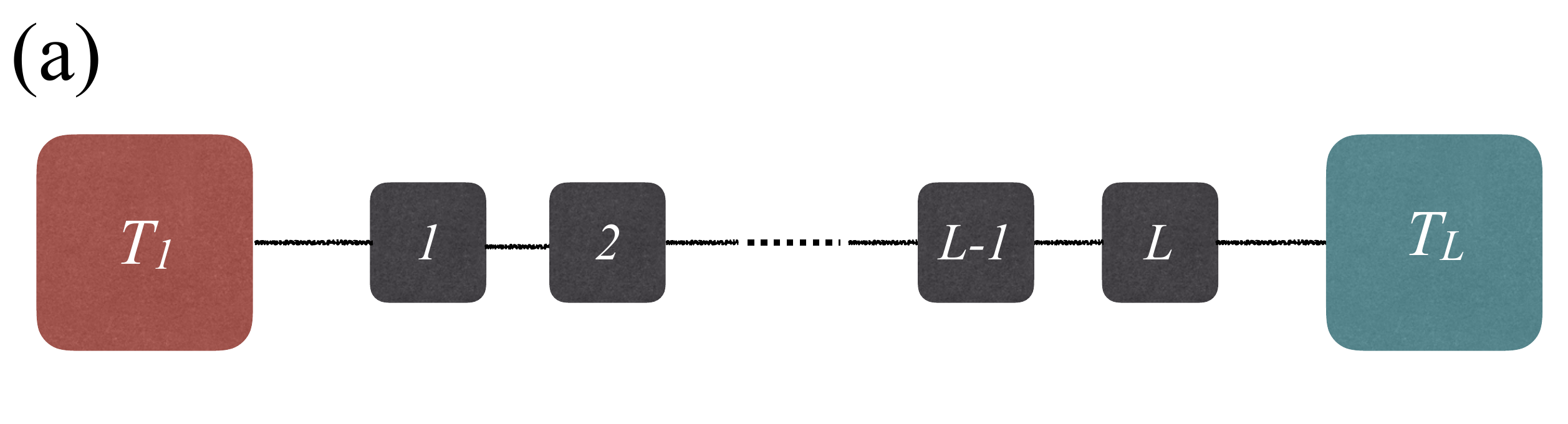}\\
\includegraphics[width=0.4\textwidth]{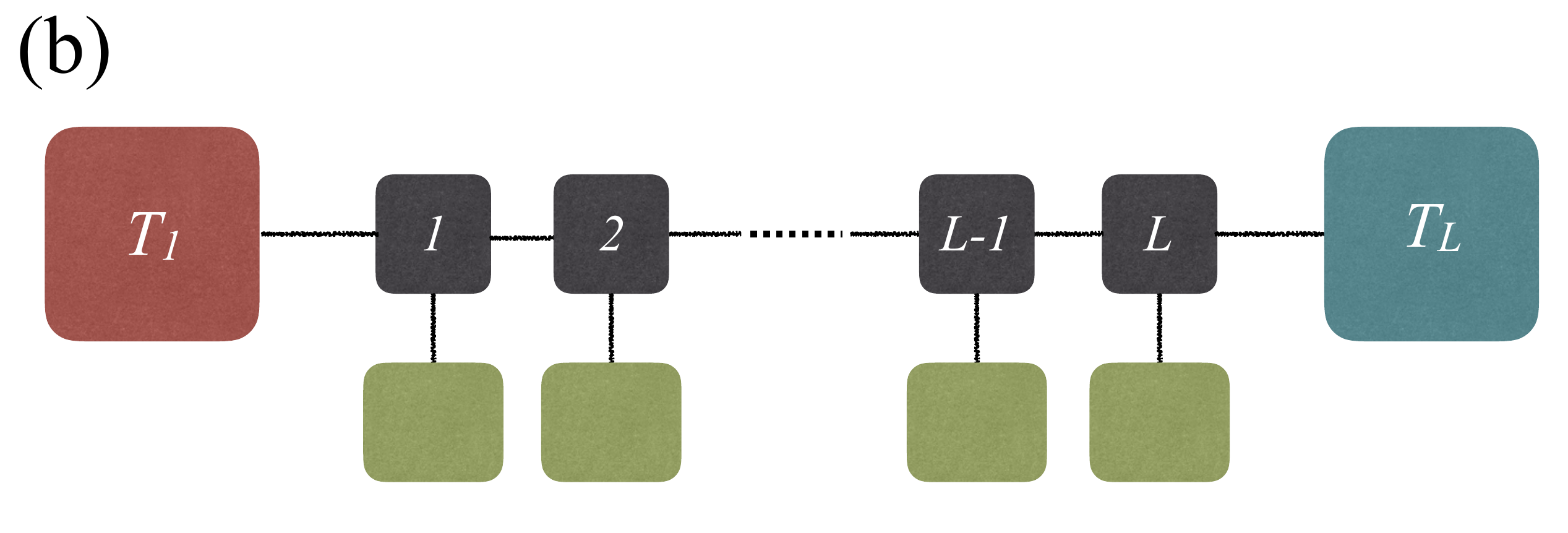}
\caption{\label{fig:drawings}
(a) A boundary-driven one-dimensional quantum chain subject to two baths at each end. 
(b) By including additional noise sources within the chain, here implemented using the idea of self-consistent reservoirs, it is possible to tune the flow from ballistic to diffusive. 
}
\end{figure}

Correlations between distant parts of a chain can be neatly quantified using the Conditional Mutual Information (CMI) $\mathcal{I}(A:C|B)$, which measures the amount of information shared between $A$ and $C$ given  $B$ \cite{Ibinson2008,Brown2012,Kato2016,Fawzi,Binder2018}.
This type of quantifier is relevant when one is interested in understanding the correlations between distant parts of the chain and how these are affected by a partition in the middle. 
In this paper we carry out a detailed study on the behavior of the CMI in the NESS of a one-dimensional bosonic chain subject to local Lindblad baths at the boundaries \cite{Karevski2009,Znidaric2011b,Znidaric2010a,Asadian2013,Landi2014a,Nicacio2015} (Fig.~\ref{fig:drawings}(a)). 
In order to introduce diffusiveness, we also add a set of self-consistent reservoirs \cite{Bolsterli1970} that function as an additional noise source (Fig.~\ref{fig:drawings}(b)).

Central to our approach is the fact that self-consistent reservoirs preserve Gaussianity and therefore allow us to use tools from Gaussian Quantum Information   \cite{Holevo1999,Adesso2014,PirlaRMP}. 
This allow us to study chains with arbitrary size and therefore find, in detail, how the CMI scales with all parameters in the system. 
As an application,  we connect our results with a recent theorem by Kato and Brand\~ao on \cite{Kato2016} the thermalization of approximate quantum Markov chains. 
We show that, from knowledge of the CMI, one can show that diffusive chains tend to be locally thermal, whereas the same cannot be said for the ballistic case. 

This paper is divided as follows. 
In Sec.~\ref{sec:CMI} we introduce the CMI,  discuss its basic features and argue why it is a relevant quantifier for the case of NESSs. 
In Sec.~\ref{sec:model} we then introduce the main model we are going to study, as well as the solution for the NESS. 
The results are then presented in Sec.~\ref{sec:res} and the conclusions in Sec.~\ref{sec:conc}. 
We also include several appendices discussing solution methods and other technical aspects.

\section{\label{sec:CMI}Conditional Mutual Information}

Consider a quantum chain with $L$ sites and prepared in a certain state  $\rho_\mathcal{S}$, where $\mathcal{S} = \{1, \ldots, L\}$ denotes the corresponding set of sites.
The reduced density matrix of a given subset  $A \subset \mathcal{S}$ is defined as $\rho_A = \tr_{\mathcal{S}/A} \rho_S$, where $\mathcal{S}/A$ stands for the partial trace over all sites that are not in the set $A$. 
The total amount of information shared between two disconnected subsets $A$ and $B$ of $\mathcal{S}$,  can be quantified by the mutual information (MI) \cite{Nielsen}
\begin{equation}\label{IAB}
\mathcal{I}(A\!:\!B) = S(\rho_A) + S(\rho_B) - S(\rho_{AB}) \geq 0,
\end{equation}
where $S(\rho) = - \tr(\rho \ln \rho)$ is the von Neumann entropy. 
This quantity measures the amount of information (quantum and classical) contained in the global state $\rho_{AB}$, but absent  in the marginalized state $\rho_A \otimes \rho_B$. 
If $\rho_{AB}$ is a pure state (which is seldom the case for NESSs), this reduces to twice the entanglement entropy. 

The MI can also be extended to the multipartite scenario. For instance, the quantity 
\begin{equation}
\mathcal{I}(A\!:\!B\! : \! C) = S(\rho_A) + S(\rho_B) + S(\rho_B)  - S(\rho_{ABC}) \geq 0,
\end{equation}
measures the total amount of information that is contained in $\rho_{ABC}$, but which is absent in the marginalized state $\rho_A\otimes \rho_B\otimes \rho_C$. 
Carrying this logic all the way towards a single site, one finds the so-called total correlations (TC) \cite{Goold2015a}
\begin{equation}\label{Itot}
\mathcal{T} = \sum\limits_{i=1}^L S(\rho_i) - S(\rho_\mathcal{S}) \geq 0 ,
\end{equation}
where $\rho_i$ is the reduced density matrix of site $i$. 
This quantity is useful in understanding correlations from a global perspective, without  reference to the geometry of the lattice or any partitioning. 

When the chain $\mathcal{S}$ is split into a bipartition of the form $A = \{1,\ldots, k\}$ and $B = \{k+1,\ldots, L\}$,  the MI~(\ref{IAB}) will fully capture the information shared between $A$ and $B$.
However, when one wishes to study disconnected bipartitions, a subtlety arises. 
Consider for instance a tripartition  $ABC$, where $A = \{1,\ldots, k\}$, $B = \{k+1,\ldots,k+b\}$ and $C = \{k+b+1,\ldots,L\}$.
While the correlations between $A$ and $C$ can also be quantified by the MI~(\ref{IAB}), this information will in general depend on the amount of knowledge one has about $B$.
This is a consequence of the fact that part of the correlations between $A$ and $C$ could be solely due to the lack of information about $B$ that both share. 
To account for this one introduces the Conditional Mutual Information (CMI) \cite{Wyner1978,Renner2002}
\begin{equation}\label{IABC}
\mathcal{I}(A\!:\!C|B) = S(\rho_{AB}) + S(\rho_{BC}) - S(\rho_{ABC}) - S(\rho_B) \geq 0.
\end{equation}
It quantifies the amount of correlations shared between $A$ and $C$, \emph{given} $B$. 
Knowledge on how $\mathcal{I}(A\!:\!C|B)$ depends on the size $b = |B|$ of the middle partition will therefore provide information on how $B$ degrades the correlations between $A$ and $C$. 
We also mention in passing that the non-negativity of the CMI is a consequence of the famous strong sub-additivity inequality \cite{Lieb1973}, which is perhaps the most important property of the von Neumann entropy.

\begin{figure}[!t]
\centering
\includegraphics[width=0.4\textwidth]{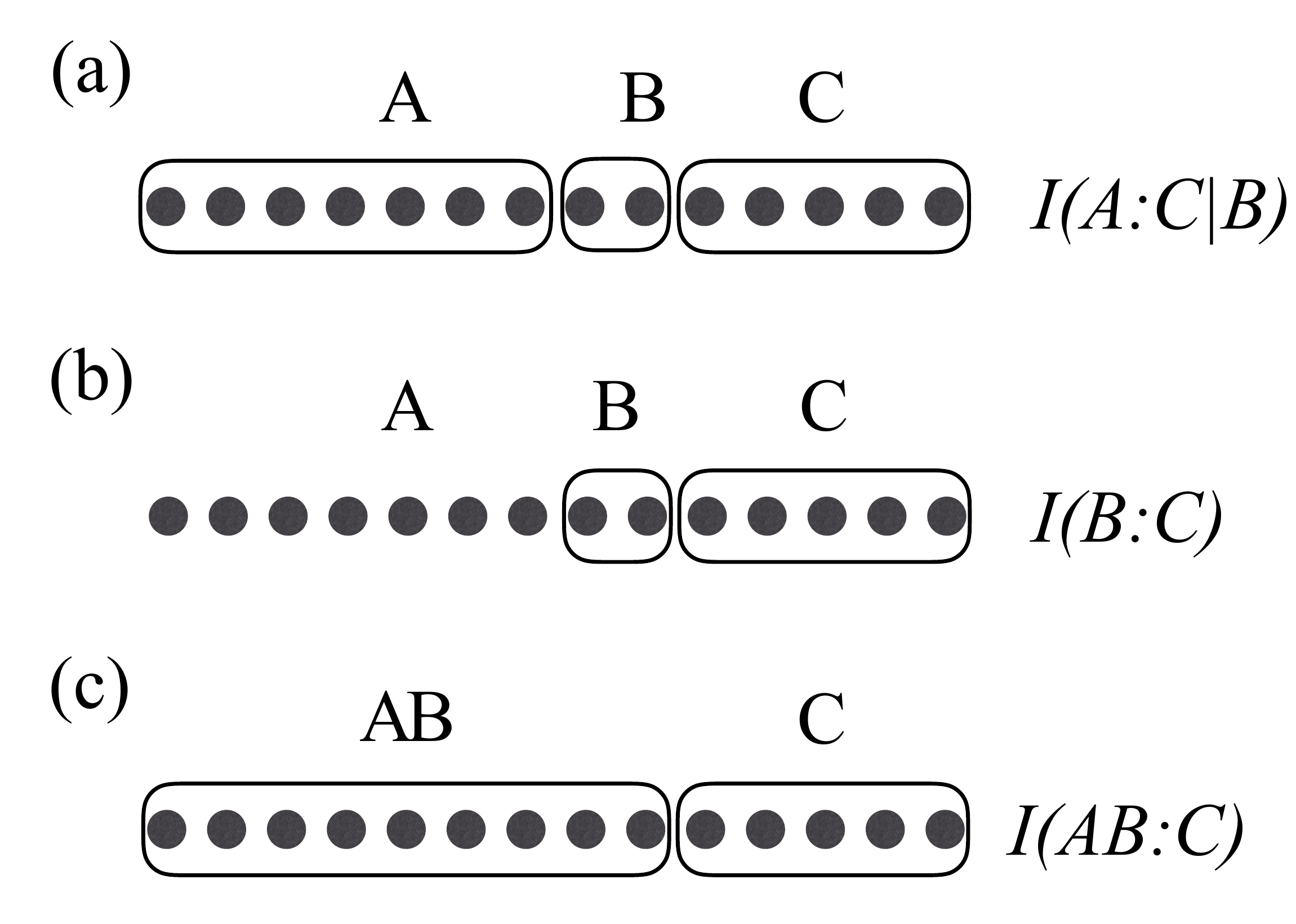}
\caption{\label{fig:diagrams_CMI}
The chain rule~(\ref{chain_rule}) connecting  the MI~(\ref{IAB}) and the CMI~(\ref{IABC}).
}
\end{figure}

The connection between the CMI~(\ref{IABC}) and the  MI~(\ref{IAB}) can be established by means of the so-called chain rule \cite{Renner2002} (see Fig.~\ref{fig:diagrams_CMI}): 
\begin{equation}\label{chain_rule}
\mathcal{I}(A\!:\!C|B) = \mathcal{I}(AB\!:\!C) - \mathcal{I}(B\!:\!C).
\end{equation}
This formula clarifies the meaning of the CMI as representing the information shared between $AB$ and $C$, subtracted from that part which refers to correlations that already exist between $B$ and $C$. 
Of course, an equivalent formula also holds with a partition the other way around. 

We now move on to study these concepts within the specific context of an exactly soluble model for a NESS, which will allow us to study the MI and CMI for arbitrary chain sizes.

\section{\label{sec:model}Boundary-driven bosonic chain under self-consistent reservoirs}

\subsection{The model }

We consider a system with  $L$ bosonic modes  $a_i$, with the same frequency $\omega$ and interacting according to the Hamiltonian 
\begin{equation}\label{H}
H = \omega \sum\limits_{i=1}^L a_i^\dagger a_i +  i \lambda \sum\limits_{i=1}^{L-1} (a_i^\dagger a_{i+1} - a_i a_{i+1}^\dagger),
\end{equation}
where $\lambda$ is a constant and the phase was chosen merely for simplicity. 
In addition, the system is also subject to two reservoirs at each end, which we choose to model by the Lindblad master equation
\begin{equation}\label{ME_ballistic}
\frac{\ud \rho}{\ud t} = -i [H,\rho] + D_1(\rho) + D_L(\rho), 
\end{equation}
where the $D_i$ are taken to be local Lindblad dissipators of the form 
\begin{equation}\label{Di}
D_i(\rho) =  (N_i+1) \mathcal{D}[a_i] +  N_i \mathcal{D}[a_i^\dagger],
\end{equation}
where $\mathcal{D}[L] = L \rho L^\dagger - \frac{1}{2} \{L^\dagger L, \rho\}$ and $N_i$ is the local Bose-Einstein thermal distribution. 
Further details about this model, including some tools from Gaussian open quantum systems, are discussed in Appendices~\ref{app:lyap} and \ref{app:local}.

The NESS of Eq.~(\ref{ME_ballistic}) can be found analytically and is known to present  a ballistic  heat flow \cite{Rieder1967,Karevski2009,Santos2016,Asadian2013}. 
Diffusive transport in general requires anharmonic interactions, which can seldom be treated analytically. 
Instead, a customary approach to induce diffusivity is to add additional energy-conserving noise sources within the chain \cite{Bolsterli1970,Dhar2011,Lukkarinen2012,Asadian2013,Landi2013a,Landi2014a,Carollo2017a}.
The typical dephasing noise used for this purpose, however, does not preserve Gaussianity. 
This, as discussed below, is essential in order to compute entropic quantities. 
Instead, we approach the problem here using the concept of self-consistent reservoirs \cite{Bolsterli1970} (called B\"uttiker probes in the quantum transport community).
For a comparison with other dephasing noises, see Appendix~\ref{app:self_consistent}).
The idea consists in adding  $L$ additional thermal reservoirs $\tilde{D}_i(\rho)$, one at each site, of the same form as Eq.~(\ref{Di}), but with temperatures chosen so as to match the local occupation number in the NESS,  $\tilde{N}_i = \langle a_i^\dagger a_i \rangle$ (Fig.~\ref{fig:drawings}(b)). 
This ensures that no current flows to the auxiliary reservoirs, but only to the physical baths at the end-points. 

Thus, instead of Eq.~(\ref{ME_ballistic}), we consider the NESS produced by the master equation, 
\begin{equation}\label{Mtotal}
\frac{\ud \rho}{\ud t} = -i [H,\rho] + \gamma D_1(\rho) + \gamma D_L(\rho) + \Gamma \sum\limits_{i=1}^L \tilde{D}_i(\rho).
\end{equation}
Here $\gamma$ is the coupling to the physical baths, whereas $\Gamma$ is the coupling to the self-consistent (auxiliary) baths. 
Hence $\Gamma$ can also be interpreted as an additional noise source responsible for changing the flow from ballistic to diffusive. 
The NESS of this model can be computed using a variation of the method used in Refs.~\cite{Karevski2009,Znidaric2011b,Znidaric2010a,Asadian2013,Landi2014a}.
We shall leave the specific details to Appendix~\ref{app:self_consistent} and focus here only on the main results. 
In Appendix~\ref{app:ness_squeezing} we also show that it is straightforward to modify this model to introduce squeezing in the reservoirs, which can be used as a tool for introducing quantum features in the NESS.

\subsection{Properties of the NESS}

\begin{figure}
\centering
\includegraphics[width=0.22\textwidth]{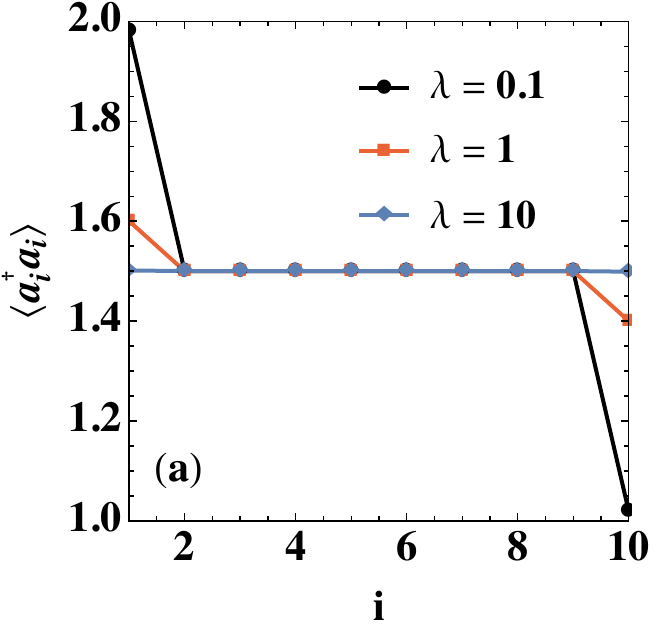}
\qquad
\includegraphics[width=0.22\textwidth]{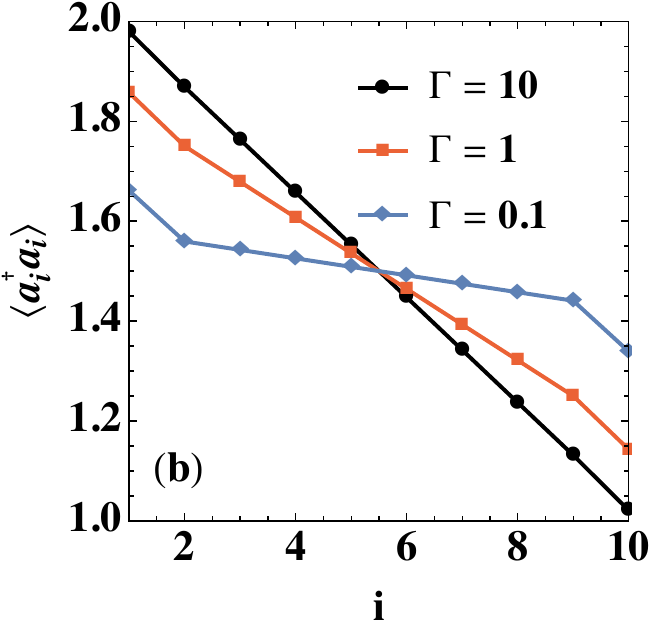}
\caption{\label{fig:SM_profile}
Occupation profile $\langle a_i^\dagger a_i\rangle$ as a function of the site number $i$.
(a) Ballistic case ($\Gamma = 0$) for different values of $\lambda$.
(b) Diffusive case, with $\lambda = 1$ and different values of $\Gamma$. 
Other parameters are $L = 10$, $\gamma  = 1$, $N_1 = 2$ and $N_L = 1$.
}
\end{figure}

The occupation numbers in the NESS, for the inner sites $i = 2, \ldots, L-1$, reads
\begin{IEEEeqnarray}{rCl}
\langle a_i^\dagger a_i \rangle &=& \frac{N_1+N_L}{2} + \frac{1}{2} \frac{\gamma  (N_1 - N_L)}{4 \lambda^2 + \gamma^2 + \gamma \Gamma (L-1)} 
 \Gamma (L-2i+1) \nonumber\\[0.2cm]
&&+ \frac{1}{2} \frac{\gamma^2  (N_1 - N_L)}{4 \lambda^2 + \gamma^2 + \gamma \Gamma (L-1)}  (\delta_{i,1} - \delta_{i,L}) .
 \IEEEeqnarraynumspace
 \label{ada}
\end{IEEEeqnarray}
For $\Gamma \equiv  0$, we obtain a flat profile typical of a  ballistic behavior,
\begin{equation}
\langle a_i^\dagger a_i \rangle = \frac{N_1+N_L}{2} + \frac{1}{2} \frac{\gamma^2  (N_1 - N_L)}{4 \lambda^2 + \gamma^2}  (\delta_{i,1} - \delta_{i,L}).
\end{equation}
This is illustrated in Fig.~\ref{fig:SM_profile}(a) for different values of $\lambda$. 
Conversely, for any $\Gamma \neq 0$ and $L$ sufficiently large, the profile approaches a linear interpolation between the bath-induced occupations $N_1$ and $N_L$,
\begin{equation}
\langle a_i^\dagger a_i \rangle \simeq \frac{(L-i) N_1 + (i-1)N_L}{L-1} + \frac{\gamma}{2\Gamma} \frac{(N_1 - N_L)}{ L-1}  (\delta_{i,1} - \delta_{i,L}).
\end{equation}
This is shown in Fig.~\ref{fig:SM_profile}(b) for different values of $\Gamma$. 
It essentially interpolates linearly between $N_1$ and $N_L$, except for small end-point corrections.

The only non-zero correlation in the system is between nearest-neighbors and reads
\begin{equation}\label{corr}
\langle a_i^\dagger a_{i+1} \rangle = \frac{\gamma \lambda}{4 \lambda^2 + \gamma^2 + \gamma \Gamma(L-1)} (N_L-N_1).
\end{equation}
This correlator determines the particle current   \cite{Karevski2009,Asadian2013}
\begin{IEEEeqnarray}{rCl}
J &=& \lambda \langle a_i^\dagger a_{i+1} + a_{i+1}^\dagger a_i \rangle 	\nonumber\\[0.2cm]
&=& \frac{2\gamma \lambda^2}{4 \lambda^2 + \gamma^2 + \gamma \Gamma(L-1)} (N_L-N_1).
\label{J}
\end{IEEEeqnarray}
Thus, we see that if $\Gamma \equiv 0$ we get a ballistic current, 
\begin{equation}
J = \frac{2\gamma\lambda^2 }{4 \lambda^2 + \gamma^2 } (N_L-N_1),
\end{equation}
which is independent of $L$.
Conversely, for any $\Gamma \neq 0$, in the limit of large $L$ we get a diffusive behavior 
\begin{equation}
J \simeq  \frac{2\lambda^2}{ \Gamma} \frac{(N_L-N_1)}{L} \sim \frac{1}{L}.
\end{equation}
A scaling of this form, with $J$ inversely proportional to $L$, is the hallmark of  Fourier's law.

\subsection{Calculation of the von Neumann entropy}

Due to the Gaussianity of the model, the state of the system is fully determined by the Covariance Matrix (CM). 
This will provide us with a simple method to compute the von Neumann entropies in terms only of symplectic eigenvalues \cite{Dutta1995,Holevo1999,PirlaRMP}
The CM of $L$ modes is defined as the $2L\times 2L$ matrix 
\begin{equation}\label{SM_CM_def}
\Theta_{i,j} = \frac{1}{2}\langle \{X_i, X_j^\dagger \} \rangle,
\end{equation}
where  $\bm{X} = (a_1,a_1^\dagger, \ldots, a_L, a_L^\dagger)$ and we have assumed the first moments are zero for simplicity. 
CMs are usually written in terms of quadrature operators  \cite{Adesso2014,PirlaRMP}.
The structure in Eq.~(\ref{SM_CM_def}), however, turns out to be quite convenient for the problem at hand since it allows us to readily separate the CM in terms of two components, 
\begin{IEEEeqnarray}{rCl}
\label{SM_C}
C_{i,j} &=& \langle a_j^\dagger a_i\rangle,
\\[0.2cm]
\label{SM_S}
B_{i,j} &=& \langle a_i a_j\rangle,
\end{IEEEeqnarray}
which are both $L\times L$.
The relation between $\Theta$ and $C$, $B$ can then be written rather elegantly as 
\begin{equation}\label{SM_Theta_SC}
\Theta = \frac{\mathbb{I}_{2L}}{2} + C\otimes (\sigma_+\sigma_-) + C\trans \otimes (\sigma_- \sigma_+) + B \otimes \sigma_+ +  B^* \otimes \sigma_-,
\end{equation}
where $\sigma_i$ are the usual Pauli matrices. 
In the NESS of Eq.~(\ref{Mtotal}), the matrix $B$ is zero, whereas $C$ is tridiagonal, with entries given by Eqs.~(\ref{ada}) and (\ref{corr}).

The von Neumann entropy for a Gaussian state can be directly computed from the symplectic eigenvalues of $\Theta$, which are defined as 
\begin{equation}
\{ \nu_k \} = \text{eigs}_+ (2 \Sigma \Theta),
\end{equation}
where $\Sigma = \mathbb{I}_L \otimes \sigma_z$ is the symplectic form related to our choice of structure for $\Theta$.
Here, $\text{eigs}_+$ means selecting only the positive eigenvalues. 
The von Neumann entropy is then given by \cite{Holevo1999,PirlaRMP}
\begin{equation}
S(\rho) = \sum\limits_{k=1}^L \bigg\{ \frac{\nu_k+1}{2} \ln \bigg(\frac{\nu_k +1 }{2} \bigg) - \frac{(\nu_k -1)}{2} \ln \bigg( \frac{\nu_k -1 }{2} \bigg)\bigg\}.
\end{equation}
The same approach is used for considering any reduced density matrices, which is also convenient since  the reduced density matrix of a Gaussian state is also Gaussian and therefore has a CM which is simply obtained by dropping from $\Theta$ the elements one wishes to trace over.

\section{\label{sec:res}Results}

\subsection{Mutual Information and Total Correlations}

\begin{figure}[!h]
\centering
\includegraphics[width=0.22\textwidth]{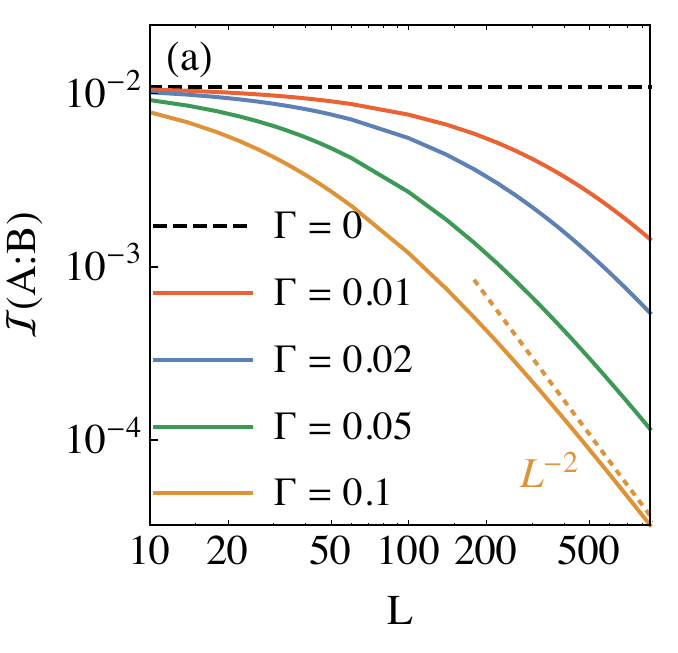}\quad
\includegraphics[width=0.22\textwidth]{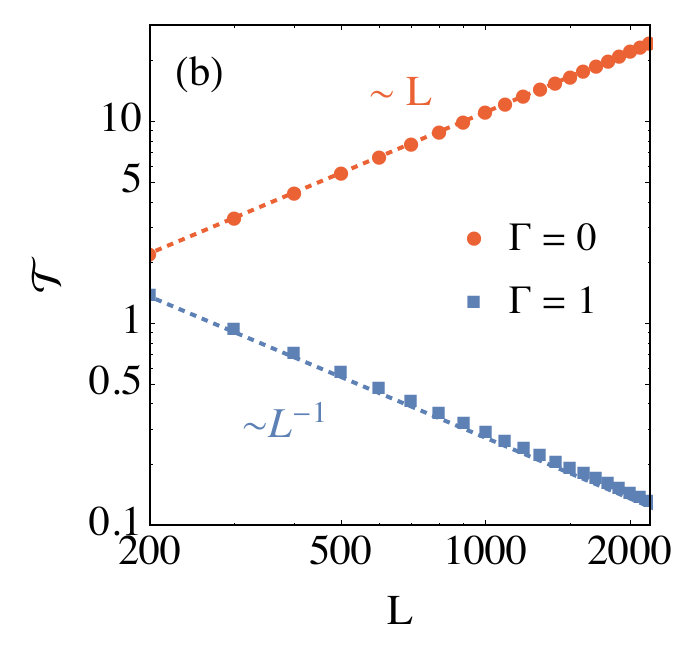}\\
\caption{\label{fig:MIT}
(a) Log-log plot of the Mutual Information (MI) $\mathcal{I}(A:B)$ [Eq.~(\ref{IAB})] between two halves of the chain as a function of $L$, for different values of the self-consistent noise $\Gamma$. 
(b) Same but for the total correlations $\mathcal{T}$ [Eq.~(\ref{Itot})]. 
The curve for $\Gamma =1$ was multiplied by $10^{-3}$ to improve visibility. 
In all curves we set $N_1 = 2$, $N_L = 1$ and $\gamma = \lambda = 1$. 
}
\end{figure}

We begin our analysis by computing the mutual information  $\mathcal{I}(A:B)$  [Eq.~(\ref{IAB})] for a symmetric bipartition at $L/2$. 
The results as a function of $L$, for different values of $\Gamma$, are shown in Fig.~\ref{fig:MIT}(a). 
As can be seen,  $\mathcal{I}(A:B)$ is independent of $L$ in the ballistic case ($\Gamma=0$), but scales as $\mathcal{I}(A:B) \sim 1/L^2$ in the diffusive case. 
In Fig.~\ref{fig:MIT}(b) we present for comparison the total correlations [Eq.~(\ref{Itot})].
For ballistic transport the TC is an extensive quantity, scaling as $\mathcal{T} \sim L$. 
Conversely, for diffusive transport we find $\mathcal{T} \sim 1/L$.
This decay of the amount of correlations as one approaches the thermodynamic limit was also found in Ref.~\cite{Znidaric2012} for 2-particle entanglement.

\subsection{Conditional Mutual Information}

\begin{figure}[!t]
\centering
\includegraphics[width=0.22\textwidth]{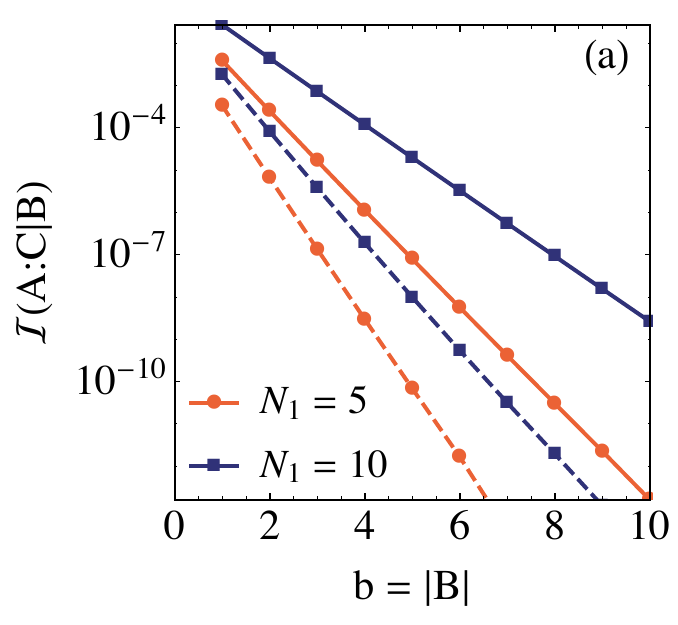}\quad
\includegraphics[width=0.22\textwidth]{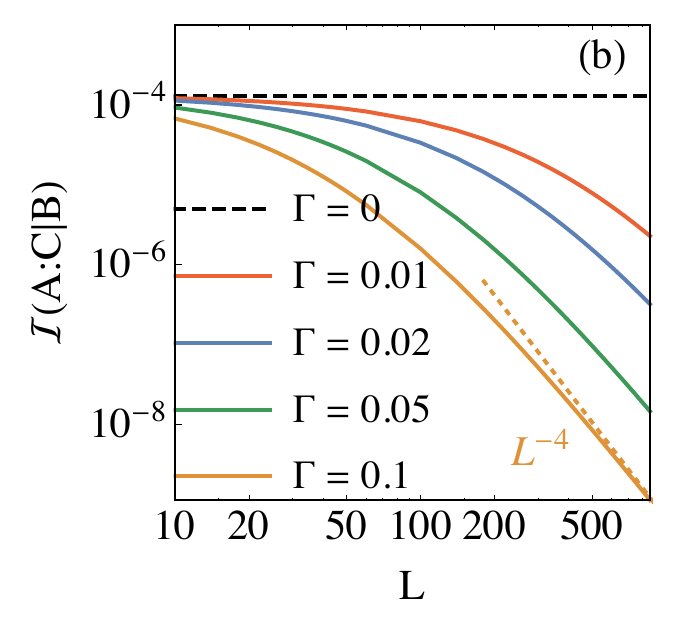}\\
\includegraphics[width=0.22\textwidth]{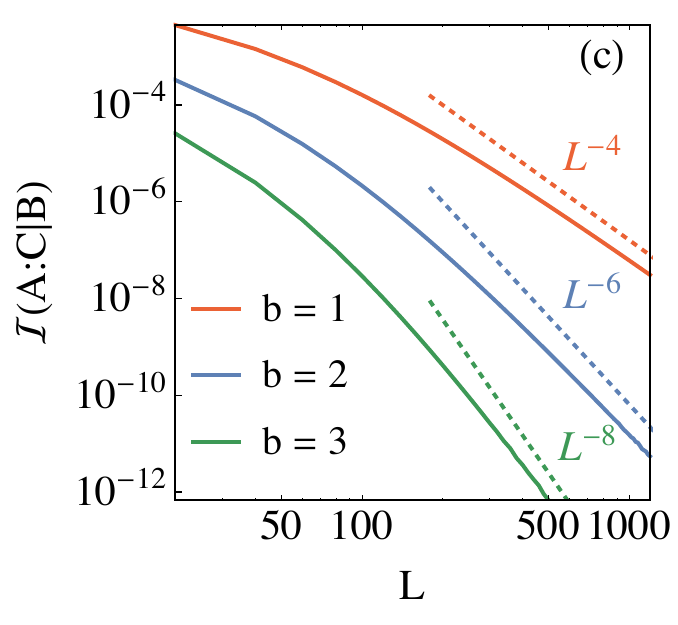}\quad
\includegraphics[width=0.22\textwidth]{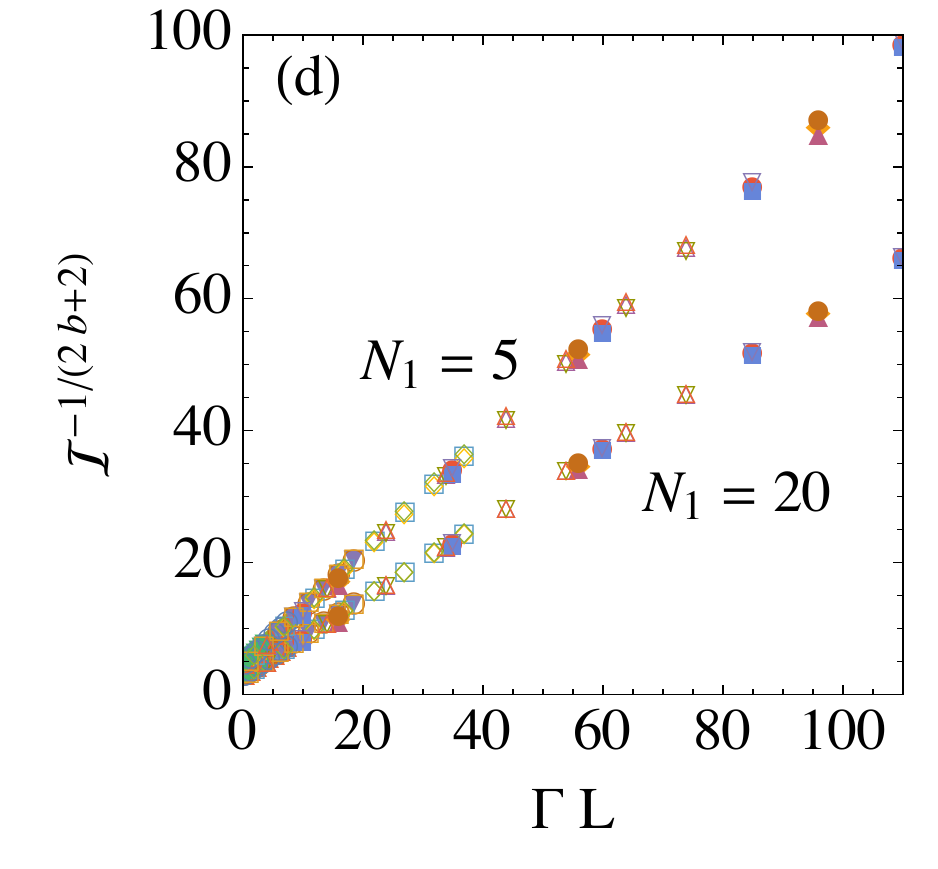}
\caption{\label{fig:CMI}
Conditional mutual information (CMI) $\mathcal{I}(A:C|B)$ [Eq.~(\ref{IABC})] for a symmetric tripartition with the middle block having size $b = |B|$.  
(a) Log of the CMI as a function of $b$, for both the ballistic case ($\Gamma = 0$, solid lines) and diffusive case ($\Gamma = 0.1$, dashed lines), with different values of $N_1$ and fixed $L = 40$ and $N_L = 1$. 
In both cases the CMI behaves as $\mathcal{I}(A:C|B) \sim 1/R^b$, where $R>1$ is a constant that depends on the temperature gradient. 
(b) Log-log plot of the CMI as a function of $L$ for $b = 1$. If $\Gamma = 0$ then $\mathcal{I}(A:C|B)$ is independent of $L$, whereas for $\Gamma \neq 0$ we get $\mathcal{I}(A:C|B)\sim 1/L^4$. 
(c) Log-log plot of  the CMI~vs.~$L$ for different values of $b$ with $N_1 = 15$, $N_L = 1$  and $\Gamma = 0.1$. 
For large $L$ the CMI scales as $\mathcal{I}(A:C|B)\sim 1/L^{2b+2}$. 
(d) Finite size scaling, Eq.~(\ref{scaling}), for two different values of $N_1$, with $N_L = 1$ and multiple values of $\Gamma$, $L$ and $b$. 
In all curves we set $\gamma = \lambda = 1$. 
}
\end{figure}

Next we turn to the CMI, which is summarized in Fig.~\ref{fig:CMI}.
We focus on symmetric tripartitions $ABC$ with $b = |B|$ sites in the middle.
As illustrated in Fig.~\ref{fig:CMI}(a), we find that in both regimes the CMI decays exponentially with  $b$, as  $\mathcal{I}(A:C|B) = 1/R^b$, 
 where $R$ is a constant that depends non-trivially on all parameters of the model. 
 The dependence of the CMI on $L$ is shown in Fig.~\ref{fig:CMI}(b) for $b=1$ and in Fig.~\ref{fig:CMI}(c) for multiple values of $b$. 
 We find that for ballistic transport  the CMI is independent of $L$, whereas for diffusive transport it scales as 
$\mathcal{I}(A:C|B) \sim 1/L^{2b+2}$.

From these  numerical simulations we therefore propose the following scaling law for the CMI:
\begin{equation}\label{scaling}
\mathcal{I}(A:C|B) = \frac{u}{(v + \Gamma L)^{2b + 2}},
\end{equation}
where $u$ and $v$ are constants. 
The behavior also holds for $b = 0$, in which case one recovers the MI in Eq.~(\ref{IAB}). 
The appearance of $L$ in the diffusive case has a dramatic consequence, as it implies that the correlations between even neighboring pats tends to zero in the thermodynamic limit when diffusivity is present. 
To confirm this scaling law behavior we present in Fig.~\ref{fig:CMI}(d)  plots of $\mathcal{I}^{-1/(2b+2)}$~vs.~$\Gamma L$ for different values of $\Gamma$, $L$ and $b$. 
According to Eq.~(\ref{scaling}), this should lead to a fully collapsed straight line, which is precisely what is observed.

Eq.~(\ref{scaling})  provides a full characterization of the information sharing for the bosonic model~(\ref{Mtotal}). 
It shows that  the information  sharing is exponentially suppressed by the size $b$ of the middle partition, for both ballistic and diffusive cases. 
However, in the diffusive case, this suppression is greatly enhanced by a factor depending on the size $L$ of the chain. 
Unfortunately, it is not possible to state whether such a scaling behavior is universal. 
It will quite likely be true for fermionic chains, in view of their similarity with the present model (c.f. Ref.~\cite{Znidaric2012}). 
There are rare situations, however, which could serve as counterexamples. 
One, for instance, is the spin helix model studied in  Ref.~\cite{Popkov2017}, where in certain cases the NESS may be in a product state, despite having a non-zero current.

\subsection{Local equilibration }

As first put forth in Ref.~\cite{Mahajan2016}, the behavior of the CMI can also shed light on questions concerning the Hilbert space tensor structure of the NESS and local equilibration. 
Motivated by this, we now show that the scaling rule~(\ref{scaling}) for the ballistic and diffusive scenarios can give precise information about how close the NESS is from local equilibrium. 
To accomplish this, we make use of a recently proved  theorem  by Kato and Brand\~ao \cite{Kato2016}.
Let $\mathcal{I}_k$ denote the CMI with a tripartition at position $k$ and only $1$ site in the middle. 
The authors have shown that if $\mathcal{I}_k < \epsilon$ for all $k$, then there exists a \emph{local} Hamiltonian $H = \sum_i h_{i,i+1}$, acting only on sites $i, i+1$, such that 
\begin{equation}\label{kato}
S\bigg(\rho_\mathcal{S} \bigg|\bigg| \frac{e^{-H}}{\tr e^{-H}}\bigg) < \epsilon L,
\end{equation}
where $S(\rho||\sigma) = \tr(\rho \ln \rho - \rho \ln \sigma)$ is the quantum relative entropy. 
This means that states with vanishingly small $\mathcal{I}_k$ tend to be \emph{locally} thermal (which includes the possibility of a site-dependent temperature, which we have incorporated into $h_{i,i+1}$).

Based on Eq.~(\ref{scaling}), with $b = 1$,  we see that in the ballistic case $\mathcal{I}_k \sim L^0$, so that the NESS will in general be far from local equilibrium. 
However, in the diffusive case $\mathcal{I}_k \sim 1/L^4$ so that Eq.~(\ref{kato}) scales as $1/L^3$. 
Hence, we see that in the diffusive case the NESS tends to a locally thermal state in the thermodynamic limit. 
This agrees with our macroscopic intuition that even though a system may be out of equilibrium it is still in a local equilibrium state, but with a position-dependent temperature. 
This result therefore provides a direct application for the CMI in understanding local properties of NESSs.

\section{\label{sec:conc}Conclusions}

In this paper we have put forth  a detailed study on the behavior of the conditional mutual information in an exactly soluble NESS. 
Our main result is summarized in Eq.~(\ref{scaling}). 
It shows that even in the ballistic case, the CMI decays exponentially with the separation $b$ between partitions $A$ and $C$. 
Moreover, in the diffusive case it also decays algebraically with the total size $L$ of the chain, hence vanishing in the thermodynamic limit. 
We also showed how this type of knowledge may find applications in studies of local thermalization of non-equilibrium states, a topic which touches at the heart of many discussions in many-body and statistical physics. 
From our studies, several natural questions emerge. 
The most basic is the universality and/or typicality of the scaling~(\ref{scaling}). 
Another interesting question is how these results would be affected by anomalous diffusion (that is, in which $J\sim 1/L^\alpha$ for some exponent $\alpha$).
By understanding what changes this would introduce in the scaling law~(\ref{scaling}), one could address the question of what is the critical value of $\alpha$ for which local equilibration breaks down.

\acknowledgements 
GTL, WTBM and GA acknowledge the S\~ao Paulo Research Foundation under grant numbers 2016/08721-7, 2017/06323-7, 2018/08211-4 and 2017/07973-5. JG is supported by an SFI-Royal Society University Research Fellowship. JG and GA acknowledge financial support from the European Research Council (ERC) under the European Union’s Horizon 2020 Research and Innovation Program (Grant Agreement Nos.~758403 and 637352).
The authors acknowledge fruitful discussions with Felix Binder,  Luis Correa and Dario Poletti.

%

%

\appendix

\section{\label{app:lyap}Lyapunov equation}

The steady-state of Gaussian bosonic problems can  be studied by solving the Lyapunov equation for the covariance matrix.
Here we shall consider a slightly more general scenario as that treated in the main text. 
In particular, we wish to show how to write down equations in which the environments also contain squeezing, as this could be useful for studying the interplay between classical and quantum information. 
 \begin{widetext}
We therefore begin by discussing the following model:
\begin{equation}\label{SM_M}
\frac{\ud \rho}{\ud t} = -i [H,\rho] + \sum\limits_{i=1}^L D_i(\rho),
\end{equation}
where $H$ is given in Eq.~(\ref{H})  and
 $D_i(\rho)$ represent local squeezed thermal baths, defined by 
\begin{IEEEeqnarray}{rCl}
\label{SM_D}
D_i(\rho) &=& \gamma_i(N_i+1)\bigg[ a_i \rho a_i^\dagger - \frac{1}{2} \{a_i^\dagger a_i, \rho\}\bigg] + \gamma_i  N_i \bigg[ a_i^\dagger \rho a_i - \frac{1}{2} \{ a_i a_i^\dagger, \rho\}\bigg]	\\[0.2cm]
 &&-\gamma_i M_i  \bigg[ a_i^\dagger \rho a_i^\dagger - \frac{1}{2} \{ a_i^\dagger a_i^\dagger, \rho\}\bigg]  - \gamma_i M_i^*   \bigg[ a_i \rho a_i - \frac{1}{2} \{ a_i a_i, \rho\}\bigg]. \nonumber
\end{IEEEeqnarray}
\end{widetext}
Here $N_i$ and $M_i$ are constants that can be associated to the thermal fluctuations and the degree of squeezing according to 
\begin{equation}\label{SM_NM}
N_i + \nicefrac{1}{2} = (\bar{n}_i + \nicefrac{1}{2} ) \cosh(2r_i), 
\qquad
M_i = (\bar{n}_i+\nicefrac{1}{2}) e^{i \theta_i} \sinh(2r_i),
\end{equation}
where $\bar{n}_i$ is the local Bose-Einstein occupation and $z_i = r_i e^{i \theta_i}$ is the local squeezing value. 
In the main text we have worked with $z_i = 0$ so that $M_i = 0$ and $N_i = \bar{n}_i$. 
Additional discussions on the use of Gaussian techniques to solve this model can also be found in Ref.~\cite{Malouf2018a}.

The dynamics of the CM under the master equation~(\ref{SM_M}) can be written as a Lyapunov equation:
\begin{equation}\label{SM_Lyapunov}
\frac{\ud \Theta}{\ud t} = \mathcal{W} \Theta + \Theta \mathcal{W}^\dagger + \mathcal{F},
\end{equation}
where
\begin{equation}\label{SM_W}
\mathcal{W}  = W \otimes \mathbb{I}_2, 
\end{equation} 
with 
\begin{equation}
W_{i,j} =  -\frac{\gamma_i}{2} \delta_{i,j} + i\omega \delta_{ij} +\lambda (\delta_{i+1,j}-\delta_{i,j+1})
\end{equation}
and
\begin{equation}\label{SM_F}
\mathcal{F} = \text{diag}(F_1,0, \ldots,0, F_L),
\end{equation}
with
\begin{equation}
F_i =\gamma_i  \begin{pmatrix}
N_i + \nicefrac{1}{2} & M_i \\[0.2cm]
M_i & N_i + \nicefrac{1}{2}
\end{pmatrix}.
\end{equation}

Instead of dealing with the full Lyapunov equation~(\ref{SM_Lyapunov}), we can use the fact that the Hamiltonian does not spontaneously generate squeezing to factor the evolution into two parts, related to the  reduced covariance matrices $C$ and $B$ in Eqs.~(\ref{SM_C}) and (\ref{SM_S}). 
Using this tensor structure in Eq.~(\ref{SM_W}) allows us to write to separate equations for $C$ and $B$:
\begin{IEEEeqnarray}{rCl}
\label{SM_eqC}
\frac{\ud C}{\ud t} &=& W C + C W^\dagger + F_N,
\\[0.2cm]
\label{SM_eqS}
\frac{\ud B}{\ud t} &=& W B + B W^\dagger + F_M,
\end{IEEEeqnarray}
where 
\begin{IEEEeqnarray}{rCl}
F_N &=& \text{diag}(\gamma_1 N_1, \ldots, \gamma_L N_L ),
\\[0.2cm]
F_M &=& \text{diag}(\gamma_1 M_1, \ldots, \gamma_L N_L  ).
\end{IEEEeqnarray}
Note how the two equations~(\ref{SM_eqC}) and (\ref{SM_eqS}) are now structurally identical,  which is a consequence of a convenient  choice of parameters in the master equation.

The NESS is then obtained by setting the right-hand side of Eqs.~(\ref{SM_eqC}) and (\ref{SM_eqS})  to zero; viz.,
\begin{IEEEeqnarray}{rCl}
\label{SM_eqC2}
W C + C W^\dagger + F_N &=& 0, \\[0.2cm]
\label{SM_eqS2}
W B + B W^\dagger + F_M &=& 0. 
\end{IEEEeqnarray}
If there is no squeezing in all environments, $z_i = 0$, then $F_M = 0$ and we therefore obtain $B = 0$. 
This is the situation considered in the main text. 

\section{\label{app:local}Local vs.~Global master equations}

In this section we clarify the physical interpretation of the master equation~(\ref{SM_M}) used in the main text. 
There is a longstanding discussion on the precise  meaning of the boundary driven (local) Lindblad dissipators used in Eq.~(\ref{SM_D}). 
As is well known, microscopic derivations of the Lindblad equation will lead to global dissipators which act non-locally on the system. 
Hence, the local forms used in Eq.~(\ref{SM_D}) are often taken as phenomenological descriptions.
However, global dissipators are also known to have serious deficiencies \cite{Wichterich2007}. 
Recently, some of us~\cite{Gonzalez2017}  have compared the local and global approaches with exact solutions based on the quantum Langevin equations, for the case of two harmonic oscillators. 
It was found that the local approach of Eq.~(\ref{SM_D}) can actually outperform the global approach depending on the parameters of the model. 

Consequently,  two important questions naturally emerge. 
The first is what is the precise meaning of the parameters $\bar{n}_1$ and $\bar{n}_L$ in Eq.~(\ref{SM_NM}). 
The results in Ref.~\cite{Gonzalez2017} show that, within certain parameter ranges, it is possible to attribute to $\bar{n}_i$ a function of the temperature of the reservoirs as 
\begin{equation}
\bar{n}_i = \frac{1}{e^{\omega/T_i} - 1}, 
\end{equation}
where $\omega$ is the frequency of oscillation of the local modes. 
Thus, within certain parameters it is correct to say that $\bar{n}_1 - \bar{n}_L$ represents a gradient of temperature. 
However, within other parameter ranges this may not be as precise. 
For this reason we avoided introducing the precise notion of temperature and, instead, interpreted $\bar{n}_1 - \bar{n}_L$ as a gradient of the boundary drives. 

The second question concerns the meaning of the current $J$ appearing in Eq.~(\ref{J})
By construction, it represents the current of particles, as it is derived from a continuity equation of the local occupation number 
\begin{equation}
\frac{\ud \langle a_i^\dagger a_i \rangle}{\ud t} = J_{i-1,i} - J_{i,i+1}, \qquad i = 2, \ldots, L-1.
\end{equation}
Similar schemes could be developed to describe a current of energy. 
However, recently some of us have shown that there is also an alternative interpretation  in terms of the method of repeated interactions \cite{DeChiara2018}.
This method can be used to generate local master equations from a physically consistent model \cite{Barra2015,Strasberg2016} where the environment is described by the sequential interactions with a series of ``environmental units'',, which are then discarded after each stroke.
As we have shown, within this framework it is possible to reconcile local master equations with thermodynamics by identifying a work cost associated with turning the interactions with the units on and off. 
Once this work term is identified, it turns out (for the specific case of a bosonic chain) that the current~(\ref{J}) will also correspond to the heat current to the baths, up to a constant $\omega$. 
Thus, to summarize, if one interprets the master equation~(\ref{SM_M}) as stemming from the method of repeated interactions, the current $J$ represents both the particle and heat currents.  
If not, then $J$ is to be interpreted solely as the current of particles.

\section{\label{app:self_consistent}Self-consistent reservoirs}

We now focus on Eq.~(\ref{SM_eqC2}). 
Analogous results can be stated for Eq.~(\ref{SM_eqS2}) by simply replacing $N_i$ with $M_i$. 
We first consider the ballistic model, in which $\gamma_1 = \gamma_L = \gamma$ and $\gamma_i = 0$ otherwise. 
The matrices $W$ and $F_N$ then become
\[
W = \begin{pmatrix}
-\gamma/2 & \lambda & 0 & 0 & 0 &  \ldots & 0 & 0  \\[0.2cm]
-\lambda & 0 & \lambda & 0& 0 & \ldots & 0 & 0 \\[0.2cm]
0 & - \lambda & 0 & \lambda & 0 & \ldots & 0 & 0 \\[0.2cm]
\vdots & \vdots& \vdots  & \vdots & \vdots &  \ddots & 0 &  0  	\\[0.2cm]
0 & 0 & 0  & 0 & 0  &\ldots & 0 &   \lambda	\\[0.2cm]
0 & 0 & 0 & 0 & 0 & \ldots &   -\lambda & -\gamma/2
\end{pmatrix},
\]
and
\[
F_N = \gamma \text{diag} (N_1, 0, \ldots, 0, N_L).
\]

We next add the self-consistent reservoirs. 
This means we should add to Eq.~(\ref{SM_eqC2}) the additional terms
\begin{equation}\label{SM_eqC3}
W C + C W^\dagger + F_N - \Gamma C + \tilde{F}_N = 0
\end{equation}
where $\Gamma$ is the coupling constant to the self-consistent reservoirs and
\[
\tilde{F}_N = \Gamma \text{diag}(\tilde{N}_1, \tilde{N}_2, \ldots, \tilde{N}_{L-1}, \tilde{N}_L).
\]
Here $\tilde{N}_i$ are the thermal occupations of the self-consistent reservoirs, which are chosen as
\[
\tilde{N}_i = \langle a_i^\dagger a_i \rangle = C_{i,i}.
\]
Thus, Eq.~(\ref{SM_eqC3}) can be written as 
\begin{equation}\label{SM_eqC4}
W C + C W^\dagger + F_N - \Gamma \Delta(C) = 0,
\end{equation}
where $\Delta(C)$ is the operation of removing all diagonals from $C$:
\[
\Delta(C) = C - \text{diag}(C_{1,1},C_{2,2}, \ldots, C_{L,L}).
\]

Eq.~(\ref{SM_eqC4}) is now formally identical to the model studied in Refs.~\cite{Znidaric2010a,Asadian2013}, which instead of using self-consistent baths, used dephasing baths of the form 
\begin{equation}\label{SM_dephasing}
\mathcal{K}(\rho) = \frac{\Gamma}{2} \sum\limits_{i=1}^L \bigg[ a_i^\dagger a_i \rho a_i^\dagger a_i - \frac{1}{2} \{(a_i^\dagger a_i)^2,\rho\} \bigg],
\end{equation}
Even though both models lead to the same equation for the covariances, it turns out that the steady-states themselves are different. 
The reason is that the dephasing model~(\ref{SM_dephasing}) does not preserve Gaussianity since the Lindblad generators are quadratic, instead of linear, in the creation and annihilation operators. 
Consequently, the NESS of the self-consistent reservoirs is Gaussian, but that of the dephasing model is not. 
This will be further discussed in the next section. 
We also mention that the self-consistent model can be readily extended for the case of squeezing, whereas the dephasing model cannot, since a dissipator such as~(\ref{SM_dephasing}) preserves the number of particles, but does not preserve the level of squeezing. 

\begin{widetext}
The solution of Eq.~(\ref{SM_eqC4}) is then identical to that studied in Ref.~\cite{Asadian2013,Znidaric2010a}. 
The matrix $C$ is tridiagonal, of the form 
\begin{equation}
C = 
\begin{pmatrix}
A_1 & x & 0 & 0 & 0 &  \ldots & 0 & 0  \\[0.2cm]
x & A_2 & x & 0& 0 & \ldots & 0 & 0 \\[0.2cm]
0 & x & A_3 & x & 0 & \ldots & 0 & 0 \\[0.2cm]
\vdots & \vdots& \vdots  & \vdots & \vdots &  \ddots & 0 &  0  	\\[0.2cm]
0 & 0 & 0  & 0 & 0  &\ldots & 0 &   x	\\[0.2cm]
0 & 0 & 0 & 0 & 0 & \ldots &   x & A_L
\end{pmatrix},
\end{equation}
where $A_i = \langle a_i^\dagger a_i\rangle$ is given in Eq.~(\ref{ada}) and $x = \langle a_i^\dagger a_{i+1} \rangle$ is given in Eq.~(\ref{corr}).

\section{\label{app:ness_squeezing}General NESS covariance matrix with squeezing}

If we introduce squeezing in the reservoirs, then Eq.~(\ref{SM_eqS2}) will give a non-trivial solution for $S$. 
This solution is identical to that for $C$, with $N_i$ replaced by $M_i$. 
Thus, in the NESS with squeezing, $S$ would also be tridiagonal, with
\begin{IEEEeqnarray}{rCl}
 R_i := \langle a_i a_i \rangle = &=& \frac{M_1+M_L}{2} + \frac{1}{2} \frac{\gamma  (M_1 - M_L)}{4 \lambda^2 + \gamma^2 + \gamma \Gamma (L-1)} 
 \Gamma (L-2i+1) + \frac{1}{2} \frac{\gamma^2  (M_1 - M_L)}{4 \lambda^2 + \gamma^2 + \gamma \Gamma (L-1)}  (\delta_{i,1} - \delta_{i,L}) ,
 \\[0.4cm]
y:= \langle a_{i} a_{i+1} \rangle  &=&
 \frac{\gamma \lambda}{4 \lambda^2 + \gamma^2 + \gamma \Gamma(L-1)} (M_L-M_1). 
\end{IEEEeqnarray}
From $C$ and $S$ we can then reconstruct the full CM $\Theta$ using Eq.~(\ref{SM_Theta_SC}). 
We then find that $\Theta$ will be block tridiagonal, of the form
\begin{equation}
\Theta = 
\begin{pmatrix}
Q_1 & Z & 0 & 0 & 0 &  \ldots & 0 & 0  \\[0.2cm]
Z & Q_2 & Z & 0& 0 & \ldots & 0 & 0 \\[0.2cm]
0 & Z & Q_3 & Z & 0 & \ldots & 0 & 0 \\[0.2cm]
\vdots & \vdots& \vdots  & \vdots & \vdots &  \ddots & 0 &  0  	\\[0.2cm]
0 & 0 & 0  & 0 & 0  &\ldots & 0 &   Z	\\[0.2cm]
0 & 0 & 0 & 0 & 0 & \ldots &   Z & Q_L
\end{pmatrix},
\end{equation}
where 
\begin{equation}
Q_i = \begin{pmatrix}
A_i +\nicefrac{1}{2}& R_i \\[0.2cm]
R_i^* & A_i +\nicefrac{1}{2}
\end{pmatrix},
\qquad\qquad
Z = \begin{pmatrix} x & y \\[0.2cm] y^* & x \end{pmatrix}.
\end{equation}

\end{widetext}

\bibliography{/Users/gtlandi/Documents/library}
\end{document}